\documentclass[12pt]{iopart} % this line compiles with larger font text
  \expandafter\let\csname equation*\endcsname\relax
  \expandafter\let\csname endequation*\endcsname\relax 
\usepackage{amsmath}
\usepackage{amsfonts}
\usepackage{amssymb}
\usepackage{bbold}
\usepackage[latin1]{inputenc}
\usepackage{graphicx}

\begin{document}

\title{The Role of Nonlinear Dynamics in Quantitative Atomic Force Microscopy}

\author{Daniel Platz, Daniel Forchheimer, Erik A. Thol{\'e}n and David B. Haviland}

\address{Nanostructure Physics, Royal Institute of Technology (KTH), Roslagstullsbacken 21, Stockholm, Sweden}

\ead{haviland@kth.se}

\begin{abstract}

Various methods of force measurement with the Atomic Force Microscope (AFM) are compared for their ability to accurately determine the tip-surface force from analysis of the nonlinear cantilever motion.  It is explained how intermodulation, or the frequency mixing of multiple drive tones by the nonlinear tip-surface force, can be used to concentrate the nonlinear motion in a narrow band of frequency near the cantilevers fundamental resonance, where accuracy and sensitivity of force measurement are greatest.  Two different methods for reconstructing tip-surface forces from intermodulation spectra are explained.  The reconstruction of both conservative and dissipative tip-surface interactions from intermodulation spectra are demonstrated on simulated data.

\end{abstract}

\maketitle

\section{Introduction}

The accurate measurement of tip-surface force is central to the development of AFM toward a more quantitative form of microscopy.  Determining tip-surface force from steady-state cantilever dynamics, as opposed to static bending, promises greater sensitivity, higher accuracy, and the ability to probe the viscous response of surfaces and other dissipative tip-surface interactions.   We see the advancement of quantitative dynamic AFM as resting on the solution to three underlying problems:  The first problem is the accurate measurement of cantilever motion, where calibration is the central issue.  The second problem is the analysis of the measured motion to correctly determine the tip-surface force.   This second problem involves nonlinear dynamics and a clear understanding of the cantilever as a transducer of tip-surface force.  The third problem is the interpretation of the properly measured tip-surface force so as to provide useful, quantitative information about the material system studied.  In this paper we discuss these three problems, with particular emphasis on the second.

We compare different modes of AFM force measurement in terms of their sensitivity and the ability to accurately calibrate the measurement.  For this purpose it is useful to analyse the cantilever motion in the frequency domain.  We do this by simulating the motion of an AFM cantilever using a point-mass, simple harmonic oscillator model, subject to a nonlinear tip-surface force.   With modern numerical integrators we can accurately determine  the motion of our model nonlinear dynamical system with the nonlinear tip-surface force is an input to the simulator.  Various measurement scenarios and drive schemes can then be compared by examining the frequency content of the motion and its relation to the noise in typical AFM experiments.  Such simulations are vital to understanding and improving AFM and they allow us to verify methods for reconstructing force from nonlinear motion.  With simulated data we can work backward from the nonlinear oscillation to a {\em known force}.   This is in contrast to experiment, where we do not know the actual nonlinear tip-surface force which contributed to the measured motion.   

Before we begin our discussion of the different dynamic force measurement methods, we will make a quick review of quasi-static force measurement which is most common among AFM users. Quasi-static and dynamic force measurements both use the same basic AFM apparatus and it is possible to compare them within the context of one simple model.  After considering the problem of reconstructing tip-surface force for both quasi-static measurement and the so-called  fast-force-curve measurement, we will consider the case of driving the cantilever near a resonance.  We describe a method called Intermodulation AFM that exploits frequency mixing by a nonlinear oscillator.  The Intermodulation method allows one to reconstruct the tip-surface force by collecting response only in a narrow frequency band near resonance, where accuracy and sensitivity are greatest.   It is explained how one can reconstruct both the conservative and dissipative forces between the tip and the surface from analysis of the intermodulation spectrum.

We have attempted to write this article in a pedagogical fashion with the hope that it will be useful to the growing community of scientists and engineers who both development and use methods of multi-frequency AFM.  A mathematical notation for preforming matrix algebra on discrete spectra is developed, which is particularly well suited for the analysis of weakly nonlinear oscillations.  While this paper deals explicitly with dynamic AFM, we would like to stress that the techniques developed herein are generally applicable to the many types of measurements that exploit the enhanced sensitivity of a transducer with a high quality resonance.  We hope that our explanation of the intermodulation spectral measurement technique and the methods for analysis of the intermodulation spectra, will lead to more wide spread application of intermodulation spectroscopy in nanotechnology.    

\section{The point mass, single eigenmode model}\label{sec:single_mode}

\begin{figure}[t]
\includegraphics[width=12cm]{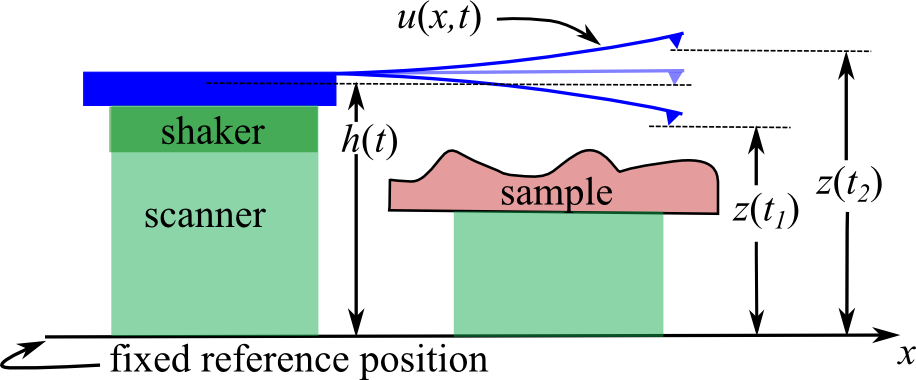}
\caption{A simplified schematic diagram of a probe-scanning AFM which explains the coordinate convention used in the text.}
\label{coordinate_convention}
\end{figure}

Let us analyse the problem of tip motion in a coordinate system fixed to the inertial reference frame in which the sample is at rest, the so-called laboratory frame.  In this coordinate system the vertical location of the tip is denoted $z$ and the vertical location of the tip when the cantilever is in its equilibrium position is denoted $h$ (see fig.~\ref{coordinate_convention}).  
We will analyse the scaned-probe AFM, where the position $h(t)$ can be changed by moving the base of the AFM probe, either rapidly with a shaker piezo, or more slowly with the Z piezo in the AFM scanner.  To model dynamic AFM, one typically reduces the spatially distributed dynamic bending of the cantilever to an equivalent point-mass model for the tip coordinate $z(t)$.   Following this approach, we postulate that the cantilever exerts a vertical force on this point-mass, such that $\mathrm{f_{c}}=-k_c (z-h)$.  We apply Newtons second law to this effective point-mass in the laboratory frame and write down an equation of motion for the tip including an inertial term, a damping term, a cantilever restoring force, and nonlinear tip-surface force $\mathrm{f}(z)$.
\begin{equation}
m \ddot{z} =  - m \gamma_0 \dot{z} - k_c (z-h) + \mathrm{f}(z)
\label{equ_mot}
\end{equation}
Here the dots mean differentiation with respect to time and the damping parameter $\gamma_0$ is interpreted as the rate of momentum loss due to random collisions with molecules of air or fluid.    In terms of the deflection $d(t)=z(t)-h(t)$ , the equation of motion becomes.  
\begin{equation}
\frac{1}{\omega_0^2} \ddot{d} + \frac{1}{Q \omega_0} \dot{d} + d = \left[ - \frac{1}{\omega_0^2} \ddot{h} - \frac{1}{Q \omega_0} \dot{h} \right] + \frac{1}{k_{\rm c}}\mathrm{f}(h + d)
\label{equ_mot_2}
\end{equation}
where $\omega_0=\sqrt{k_c/m}$ is the resonant frequency and $Q=\omega_0/\gamma_0$ is the quality factor.  

The equivalent point-mass model eq.~(\ref{equ_mot_2}) is only approximate, but it does capture the basic features of AFM cantilever dynamics under many realistic conditions of operation. In fact, under certain conditions of operation,  eq.~(\ref{equ_mot_2}) is actually very accurate, as will be discussed below.  The two terms in square brackets are often referred to as the "drive force".  The first is an inertial force that arises because the cantilever has mass and we are shaking (accelerating) the base, and the second is a drag force due to the motion of the cantilever through the damping medium.    In what follows, we will investigate the nonlinear dynamics of a cantilever driven by a moving base and interacting with a surface, by numerical integration of eq.~(\ref{equ_mot_2}).  

A rigorous derivation of the equivalent point-mass model is a subtle and difficult problem that involves solving the Euler-Bernouli equation governing the beam dynamics in the non-intertial reference frame attached to the base \cite{Lee:NonLinDynamMicrolever:02}\cite{Rodriguez:PointMassHarmonics:02}\cite{Melcher:EquivPointMassModel:07} which is moving through a damping medium \cite{Sader:FreqResponseBeamViscous:98}.   The solution to the resulting equation is a formidable task which we do not discuss here.  We only note that the  equation is separable in space and time, so that a particular solution for the dynamic deflection of the beam in the moving frame can be written as $u(x,t)=\Phi(x)q(t)$.   The vibrational eigenmodes of the ideal, undamped beam, rigidly fixed at one end and free at the other end, constituent an orthogonal basis set of bending functions $\Phi_i(x)$ with which we could describe an arbitrary deflection of the free end of the beam.  The time dependence of the ideal beam is described by a harmonic oscillator equation with a specific eigenfrequency for each bending function.  These eigenmodes are coupled by the damping force, which is a function of $x$, and the non-linear tip-surface force, which acts at the free end of the beam  $\mathrm{f}(z)=\mathrm{f}( h+\sum_i a_i\Phi_i(L_c)q_i(t))$.  Thus, the general problem is that of a coupled, multi-dimensional nonlinear dynamical system, with two dimensions for each relevant eigenmode. 

The problem of determining tip-surface force from cantilever motion in a realistic AFM experiment is therefore quite complex {\em if the motion involves many eigenmodes}.   However, the problem is simplified enormously if all significant motion is contained in a narrow frequency band surrounding the fundamental eigenfrequency.  Concentrating nonlinear motion to one resonance is possible with cantilevers because they do not have their eigenfrequencies at integer multiples of a fundamental.  This is in contrast to other common resonators such as taught strings on a musical instrument or electromagnetic transmission-line-resonators which do have their eigenfrequencies at integer multiples of a fundamental.  If the cantilever eigenmodes have high quality resonances, multiple modes will not be excited by the nonlinear tip-surface force which has frequency content at harmonics, or integer multiples of the drive frequency.  When the quality factor is too low, the frequency content of the nonlinear force will overlap with and excite multiple modes of the cantilever, as observed for example in liquids \cite{Xu:ForceTappingLiquid:10}.  Concentrating motion to the fundamental eigenmode is advantageous because the shape of this mode is particularity insensitive to the viscosity of the damping medium \cite{Sader:FreqResponseBeamViscous:98}, a fact which opens up the possibility of using thermal noise as a means of accurately calibrating all mode parameters and the deflection detector, with one simple noise measurement.  

These considerations motivate a comparison of the different methods of dynamic force measurement in terms of the frequency content of the cantilever motion.  

\section{Frequency domain analysis}\label{sec:frequency_domain}

In an AFM experiment the motion $d(t)$ is typically detected by optical means and converted to an electronic signal.  This signal is digitized by sampling at equally spaced intervals $\Delta t$ for a finite time $N \Delta t$.  We can represent the motion in this observation window in either the time domain or the frequency domain, where two are related by the Discrete Fourier transform (DFT) and its inverse.  
\begin{eqnarray}
\hat{d}_k &\equiv &
\sum_{n=0}^{N-1} d(n \Delta t) e^{-\frac{2 \pi i}{N} k n} \label{DFT}\\
d_n & \equiv & \frac{1}{N} \sum_{k=0}^{N-1}  \mathrm{\hat{d}}(k \Delta \omega) e^{+\frac{2 \pi i}{N} k n} 
\end{eqnarray}
 The time domain data $d_n$ consists of $N$ real data values, and the spectrum $\hat{d}_k=\hat{d}(k \Delta \omega)$ consists of $N/2$ complex numbers equally spaced in frequency by $\Delta \omega = 2 \pi / N \Delta t$ , where N is assumed to be even.  In what follows, we will represent these discrete functions of frequency and time as vectors denoted by lower case bold letters, where a hat is used to denote functions of frequency
\begin{eqnarray}
\mathbf{\hat{d}} &=& \mathcal{F} \left[  \mathbf{d} \right] \\
\mathbf{d} &=& \mathcal{F}^{-1} \left[  \mathbf{\hat{d}} \right] 
\end{eqnarray}

Let us first consider the frequency spectrum of the free motion within the context of the single eigenmode model.  When the cantilever is well above the surface the nonlinear tip-surface force is zero.  The DFT of equation (\ref{equ_mot_2}) with $\mathrm{f}=0$ can be written as a matrix equation for the discrete response spectrum of the free cantilever.  
\begin{equation}
\mathbf{\hat{d}}^{\rm (free)} = \left( \mathbf{\hat{G}} - \mathbb{1} \right) \mathbf{\hat{h}}
\label{delta_free}
\end{equation}
The matrix $\mathbb{1}$ is the identity matrix and the diagonal matrix $\mathbf{\hat{G}}$ is the linear transfer function of the harmonic oscillator.  
\begin{equation}
\hat{G}_{kl}=\begin{cases}
\begin{array}{c}
\left[ 1+i \left( \frac{k \Delta \omega}{\omega_0 Q} \right) - \left( \frac{k \Delta \omega}{\omega_0} \right)^2 \right]^{-1} \\ 0
\end{array} & \begin{array}{c}
k=l \\
k \neq l
\end{array}\end{cases}
\label{G}
\end{equation}
 Multiplication by the complex numbers that are the diagonal elements of $(\mathbf{\hat{G}}-\mathbb{1})$, transfers the drive spectrum $\mathbf{\hat{h}}$, which is the motion of the base, to a response spectrum $\mathbf{\hat{d}}^{\rm (free)}$, which is the free deflection of the cantilever.  The factor $\mathbb{1}$ is negligable when the quality factor is high and one is driving on or near resonance, where $\vert \mathbf{\hat{G}} \vert \simeq Q >> 1$.   On the other hand, driving at  frequencies well below a high $Q$ resonance, where $\vert \mathbf{\hat{G}} \vert \simeq 1$, generates no deflection of the free cantilever.    Note  that linearity of the response is expressed in the diagonal nature of  the transfer matrix, which maps each frequency component of the drive vector $\mathbf{\hat{h}}$ to the same frequency in the response vector $\mathbf{\hat{d}}^{\rm (free)}$.  

Consider next the single eigenmode dynamics while the cantilever is engaging a surface, where the nonlinear tip-surface force is not zero.  Taking the DFT of the equation of motion (\ref{equ_mot_2}) leads to the following equation for the frequency spectrum of the engaged cantilever deflection $\mathbf{\hat{d}}$ .  
\begin{equation}
\mathbf{\hat{d}} = \mathbf{\hat{d}}^{\rm (free)} + k_c^{-1}\mathbf{\hat{G}} \mathbf{\hat{f}}
\label{engaged_spectrum}
\end{equation}
Here we have assumed that the free spectrum and the engaged spectrum are both generated by the same drive $\mathbf{\hat{h}}$.  In doing so we are neglecting the back-action of the tip-surface force on the actuator which is providing the drive.  This is clearly allowed as the cantilever base,  as well as the shaker and scanner piezo actuators, all have far greater mass than the effective mass of the cantilever eigenmode.

Due to the nonlinear nature of the tip-surface force, the engaged spectrum $\mathbf{\hat{d}}$ will have components in the response spectrum at frequencies where there is no drive.  By measuring these frequency components, we can determine the nonlinear force, which can be seen by inverting eq.(\ref{engaged_spectrum}).
\begin{equation}
\mathbf{\hat{f}}=k_c \mathbf{\hat{G}^{-1}}(\mathbf{\hat{d}} - \mathbf{\hat{d}^{(free)}})
%=k_c \mathbf{\hat{G}^{-1}}(\mathbf{\hat{d}} +\mathbf{\hat{h}})  - \mathbf{\hat{h}}
\label{inversion_1}
\end{equation}
%Thus, by subtracting the free, linear response of the cantilever from the engaged nonlinear response, we can determine the frequency components of the nonlinear force $\mathbf{\hat{f}}$, provided that we know the inverse of the linear transfer function $\mathbf{\hat{G}^{-1}}$.  

Equation (\ref{inversion_1}) gives us the Fourier components of the tip-surface force, but we are actually interested in the tip-surface force as a function of the tip position $z$.  We can find $\mathrm{f}(t)$ by applying the inverse Fourier transform to eq. (\ref{inversion_1}). If the time dependence of the tip-surface force is implicit, coming only from time dependence of the tip position, the force curve  $\mathrm{f} \left( z(t) \right)$ is found by plotting $\mathrm{f}(t)$ versus $ z(t)=d(t)+h(t) $.  Note that our assumption of implicit time dependence is only valid for a conservative tip-surface force.   Non-conservative forces will be discussed in section \ref{sec:conservative_dissipative}.

%With this theoretical framework we can begin to compare various methods of dynamic AFM, and discuss where inaccuracies in the determination of $\mathrm{f}(z)$ may arise.  

\section{Sensitivity and Calibration}\label{sec:Sensitivity_Calibration}

Quantitative AFM must not only strive to fully understand the cantilever motion and its relation to all forces involved, but also the accurate calibration of the constants that are necessary for converting motion to force, namely $k$, $m$, and $\gamma$ for all relevant eigenmodes.  Furthermore,  quantitative AFM should strive to maximize the sensitivity of measurement.  Naively, one might think that sensitivity is enhanced when using a softer cantilever (smaller $k$).  It is true that the cantilever response (deflection of the free end) to a {\em static} tip-surface force increases with decreasing stiffness.   However, when we take into account the cantilevers high quality factor resonance, we come to a different conclusion when considering the frequency dependence of the cantilevers response to a {\em dynamic} force.  

The sensitivity of a measurement is expressed in terms of the signal-to-noise ratio (SNR).  Within the context of the single eigenmode model, the frequency dependant SNR is given by,
\begin{equation}
\mathrm{SNR} = \frac{  k_c^{-1}  \mathbf{\hat{G}} \mathbf{\hat{f}} }{ 
\sqrt{ k_c^{-2} \vert  \mathbf{\hat{G}} \vert ^2  S_\mathrm{FF} B   + \alpha ^2 S_{VV}   B    }}
\label{SNR}
\end{equation}
where $B$ [Hz] is the measurement bandwidth,  $S_\mathrm{FF}$  [N$^2$/Hz] and $S_{VV}$ [V$^2/$Hz] are the power spectral densities of force and detector voltage fluctuations respectively, and $\alpha$ [nm/V] is the detector responsivity.  The numerator (signal) is the frequency-dependant deflection caused by each frequency component of the tip-surface force.  The denominator (noise) consists of two independent contributions, where the square amplitudes are added to get the total noise.  The first noise term is the frequency-dependant cantilever response to the frequency-independent thermal noise force as given by the fluctuation dissipation theorem, $ S_\mathrm{FF}= 2 \mathrm{k_B} T m \gamma_0  $.  The second noise term is the detector's voltage noise, which is taken to be frequency-independent in the band of interest, expressed as an equivalent cantilever deflection noise.

\begin{figure}[t]
\begin{center}
\includegraphics[width=12cm]{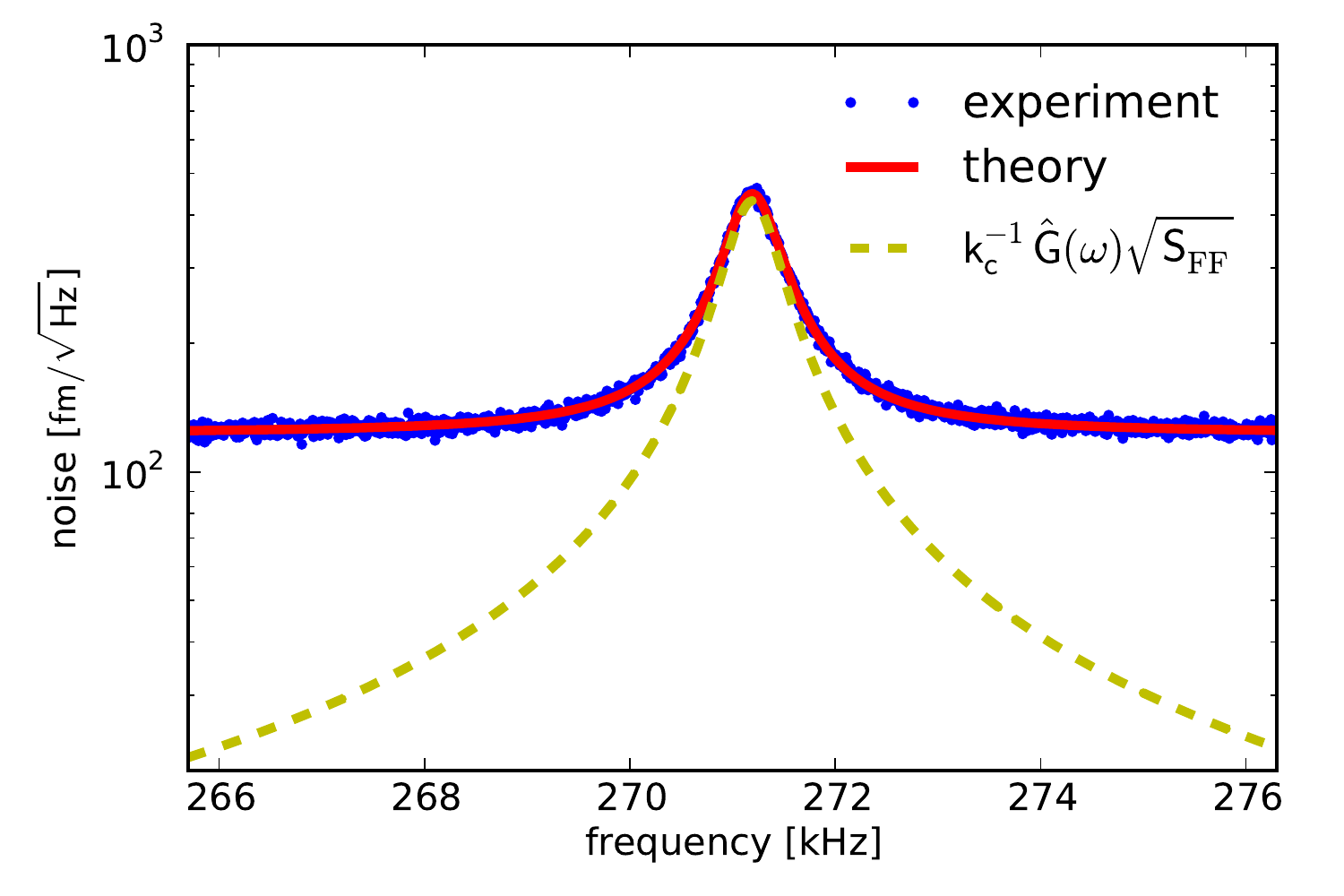}
\caption{A measurement of the noise in an AFM when the cantilever is far above a surface.  In the narrow band of frequency around the fundamental eigenfrequency, the thermal noise force can be resolved above the detector noise due to large transfer gain $G(\omega)$ of the high $Q$ resonance.  In this frequency band force measurement is most sensitive, limited only by the thermal noise force associated with the damping medium.  From the noise measurement we can calibrate all parameters in the single eigenmode model, and the optical detector responsivity $\alpha$.  In room temperature air, this cantilever had $k_c=25.8~\mathrm{N/m}$, $f_0=271~\mathrm{kHz}$ and $Q=500$.  The thermal noise force was $\sqrt{S_\mathrm{FF}}=23~\mathrm{fN/\sqrt{Hz}}$ and the optical detector noise floor was $\alpha \sqrt{S_{VV}}=120~\mathrm{fm/\sqrt{Hz}}$ . }
\label{cantilever_noise}
\end{center}
\end{figure}

Figure ~\ref{cantilever_noise}  shows a fit of the theory (denominator of eq.~(\ref{SNR})) to the measured noise for an undriven cantilever far above a surface, where there are no tip surface forces.  Here we display actual experimental data (all other figures are simulations)  taken at room temperature in air for an AFM cantilever having nominal  $f_0=300$~kHz and $k_c=40$~N/m.  The optical detection system was that of a Digital Instruments AFM and the cantilever was back side coated for enahnced reflectivity.  The theory gives an excellent description of the measurement in the band of frequency near the cantilever resonance.  

In the frequency band where the thermal noise force dominates over the noise of the detector, the SNR is independent of the cantilever stiffness $k_c$.   In this band, no improvements in the detector will effect the sensitivity of force measurement.  Improved force sensitivity can only be achieved by reducing the temperature,  or by reducing the damping coefficient $m \gamma_0$, for example by using a smaller cantilever \cite{Walters:SmallCantilevers:96} or by working in vacuum.  Signals collected in this frequency band therefore represent measurement at a fundamental limit of sensitivity for a particular cantilever and given experimental conditions.  For the cantilever of fig.~\ref{cantilever_noise} this sensitivity limit in air at room temperature is 23~fN/$\sqrt{\mathrm{Hz}}$.    We can compare this dynamic force sensitivity near resonance with the force sensitivity of quasi-static, or low frequency force measurement where $\vert \hat{G} \vert =1$,  which is given by $k_c \alpha  \sqrt{S_{VV}} = 3100$~fN/$\sqrt{\mathrm{Hz}}$.  The SNR improves by a factor of 134 for signals measured near resonance.  

Fitting the measured noise to eq.~(\ref{SNR}) provides an accurate way to determine the linear response function of the single eigenmode in question.  For the fundamental eigenmode, not only the force transducers linear response function can be determined by thermal noise measurement, but also the optical lever responsivity $\alpha$ .  Recently it was shown how  calculations of fluid dynamic damping \cite{Sader:FreqResponseBeamViscous:98} can be combined with measurements of thermal noise \cite{Hutter:NoiseCalib:93} to determine both $\alpha$ and $k_c^{-1} \mathbf{\hat{G}}$ from one simple noise measurement \cite{Higins:NoninvasiveCalibration:06}.  This so-called non-invasive method is by far the best calibration method for many reasons:  It can be preformed without touching a surface, thus keeping the tip pristine; it directly applies fundamental theory in a proper regime of validity and does not require uncontrolled assumptions about tip-sliding and surface deformation;  the calibration of all relevant constants, including $\alpha$, are traceable to one and the same measurement.  In these respects, the non-invasive method is the most 'primary' of all AFM calibration methods.  The method does however assume a single eigenmode model of the cantilevers Brownian motion, and is therefore only valid for frequencies near a resonance.    Because the mode shape of higher bending modes is sensitive to the viscosity of the damping medium \cite{Sader:FreqResponseBeamViscous:98}, the application of this non-invasive calibration method to resonances other than the fundamental bending mode has proven difficult\cite{Lozano:CalibrationHigherModes:10}.  

Thus, there is significant advantage to a method that can extract the tip-surface force by analyzing the motion only in a narrow band around the fundamental resonance of the cantilever.  In this case we can accurately describe the motion in therms of the single eigenmode model, and we have a simple and direct means of calibrating the relevant constants in our equation of motion.  Before discussing such a method in detail, we will review the more common methods of force measurement for the purpose of comparison.  

%Our discussion of the SNR emphasizes why it is so advantageous to work near resonance in AFM measurements.  Indeed, the enhanced sensitivity near resonance is the main reason that the so-called tapping mode of dynamic AFM is the most widely used imaging modality in AFM.  The theory behind eq.~(\ref{SNR}) is the linear response theory of the force transducer, described by the response function $k_c^{-1}\mathbf{\hat{G}}$.  The transducer has two degrees of freedom $d$ and $\dot{d}$,  which are taken to be in thermal equilibrium at the temperature $T$ with many microscopic degrees of freedom who's random motion causes dissipation.  The good agreement between experiment and theory speaks for the validity of the single eigenmode model in the band of frequency around resonance.   The fit of the noise data to this simple theory provides a direct  method to determine the linear response function $k_c^{-1} \mathbf{\hat{G}}$ in this frequency band cite{Higgins original paper, Sader review}.  

\section{Quasi-static force measurement}\label{sec:quasi_static}

When the cantilever base is moved with a very slow linear ramp toward and away from the surface at a rate of 0.1-10 Hz, one can neglect all time derivatives of $h$ in eq.~(\ref{equ_mot_2}).  If we also neglect the time derivatives of $d$, we are left with the equation of static force balance between the cantilever and the tip-surface force.  
\begin{equation}
k_c d = \mathrm{f}(d+h)
\label{static_force_eq}
\end{equation}
Figure~\ref{static_force_curves} shows a simulation based on this equation.  In the simulattion of fig.~\ref{static_force_curves}  and throughout this article,  we use the van der Waals - DMT model \cite{garcia:DynamicAFMReview:02} to simulate the tip surface force
\begin{equation}
\mathrm{f}(z)=\begin{cases}
\begin{array}{c}
-\frac{HR}{6z^2}\\
-\frac{HR}{6a_{0}^2}+\frac{4}{3}E^{*}\sqrt{R(a_0-z)^{3}}
\end{array} & \begin{array}{c}
z \geq a_0\\
z< a_0
\end{array}\end{cases}
\label{DMT}
\end{equation}
with the parameters:  intermolecular distance $a_0=0.3$~nm, Hamaker constant $H=7.1 \times 10^{-20}$~J, effective modulus $E^*=1.0$~GPa, and tip radius $R=10$~nm.  This model is derived for an ideal geometry with materials of uniform composition.  The actual tip-surfaces forces in real experiments may differ significantly from this model, depending on the shape of the tip, the varying topography of the surface in question, the presence or absence of adsorbed molecules on the surface, and the possibility of other, non-van der Waals interactions that could attract the tip to the surface.  Nevertheless, this simple model captures some basic features of realistic tip-surface forces, and it serves as a useful starting point for exploring the nonlinear motion of the AFM cantilever.  

Under typical experimental conditions the static equilibrium becomes unstable very close to tip contact with the surface, where strong attractive forces cause a 'jump-to-contact' upon approaching the surface, and 'pull-off' event when retracting from the surface.   This instability results in vertical jumps in a plot of the measured  cantilever deflection $d$, versus the base position $h$ (left panel of fig.~\ref{static_force_curves}).   If the spring constant $k_c$ is known, the measured quasi-static deflection gives us the tip-surface force by eq.~({\ref{static_force_eq}).  If the base position $h$ is also known, we can plot this force versus the tip position $z=d+h$ as shown in the left panel of fig.~\ref{static_force_curves}.   Here we can see that the jump-to-contact and pull-off events connect two stable solutions of eq.~(\ref{static_force_eq}) along a 'load line' with slope given by the cantilever stiffness, $k_c = 1$~N/m in this simulation.  From  fig.~\ref{static_force_curves} it is apparent that the softer the cantilever (smaller slope of the load line), the larger the bistable region in the measurement.  The soft cantilever is unable to make a static force measurement on the unstable branch of the force curve, between the two jump events.   When crossing the jump points, a rapid motion of the cantilever ensues, exciting high frequency components in the response spectrum $\mathbf{\hat{d}}$.  This high frequency motion is typically filtered out electronically when measuring a static force curve.  

\begin{figure}
\begin{center}
\includegraphics[width=16cm]{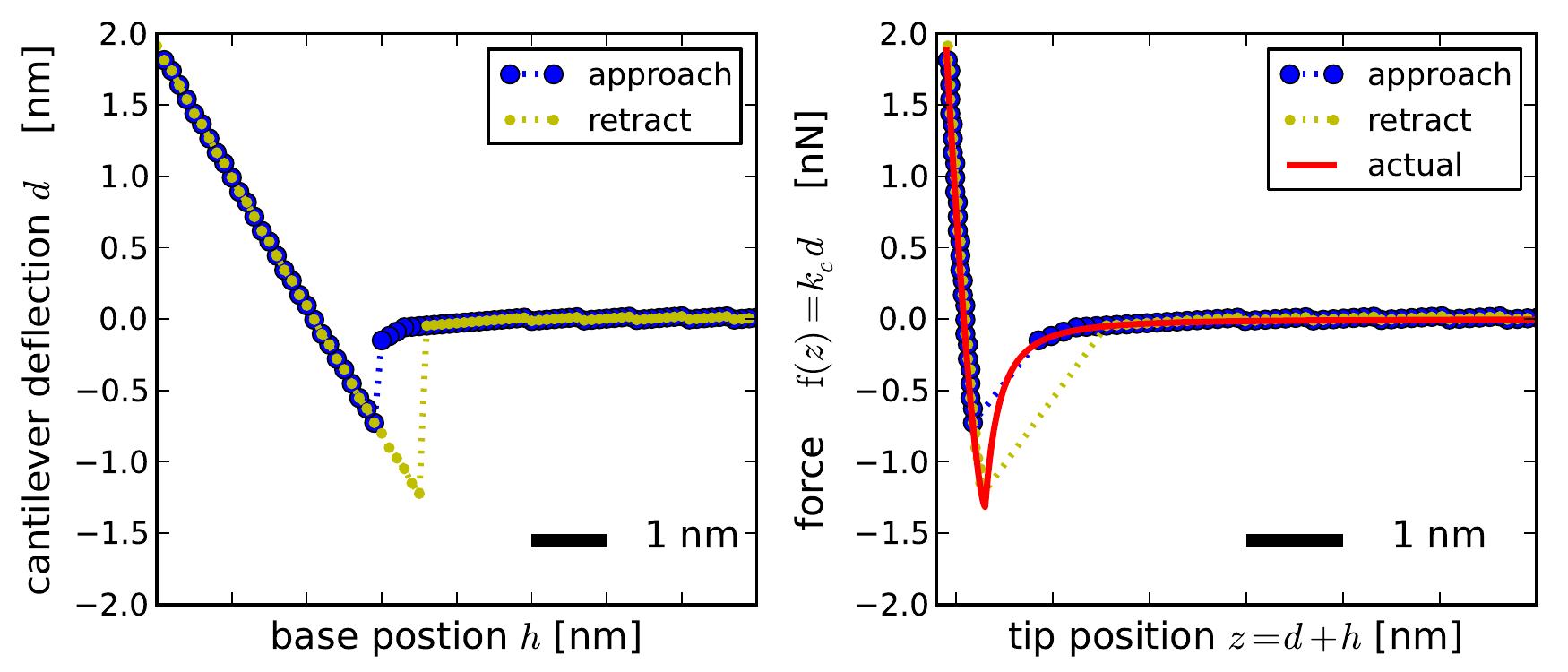}
\caption{Simulation of a quasi-static force measurement.  Left panel: The measured cantilever deflection $d$ plotted versus the position of the cantilever base $h$.  Right panel:  The tip-surface force plotted versus the tip position $z$.  The tip-surface force model used in this simulation and all subsequent simulations is described by eq.~(\ref{DMT}) with parameters given in the text.  The cantilever stiffness $k_c=1$~N/m .}
\label{static_force_curves}
\end{center}
\end{figure}

The accuracy of this quasi-static method requires that we calibrate the measurement of both $h$ and $d$.  Measuring $h$ accurately requires a well-calibrated scanner.  Measuring $d$ accurately requires a well calibrated detector.  It is advantageous if both of these measurements are traceable back to one and the same calibration.   One should be weary of systematic errors in the determination of $z=d+h$ if  the measurement of $d$ is based on thermal noise calibration,  and the measurement of $h$ relies on the scanner calibration.  The conversion of the measured quasi-static deflection $d$ to force also requires a calibration of the static force constant, which is slightly different than the dynamic mode stiffness as determined by thermal noise calibration \cite{Walters:SmallCantilevers:96}.  Finally, quasi-static force measurements suffer from the fact that the measured signals $d$ and $h$ are at very low frequencies where $1/f$ noise becomes a significant source of error.  %$1/f$ noise is ubiquitous in all physical systems, and can be thought of as a slow drifting of a measured quatity.  A drift in the measured quantities $d$ and $h$ can lead to an apparent hysteresis, or offset between the measured static force on approach and retract, which could be misinterpreted as dissipative tip-surface forces.

\section{Fast force curves}\label{sec:fast_force_curves}

The slow nature of the quasi-static method makes it impractical for high-density measurements of the tip-surface force in the x-y plane, the so-called force-volume measurement.
%\footnote{A single eigenmode is only capable of giving us the one component of force which is acting along the path of the oscillating free end.  This is approximately the vertical or $z$ component of force,  so by force-volume we mean $\mathrm{f}_z(x,y,z)$.} 
There has therefore been considerable interest in increasing the speed of force measurement so that one can map the tip-surface force with high resolution.  One approach to speeding up force measurement is to simply drive the base faster with a sinusoidal motion \cite{Krotil:pulse_force:99}.  In the commercial modes that use this method, one typically drives the cantilever base with a frequency in the range 1-3 kHz, while measuring the cantilever deflection signal over a much broader frequency band.  

\begin{figure}[t]
\begin{center}
\includegraphics[width=12cm]{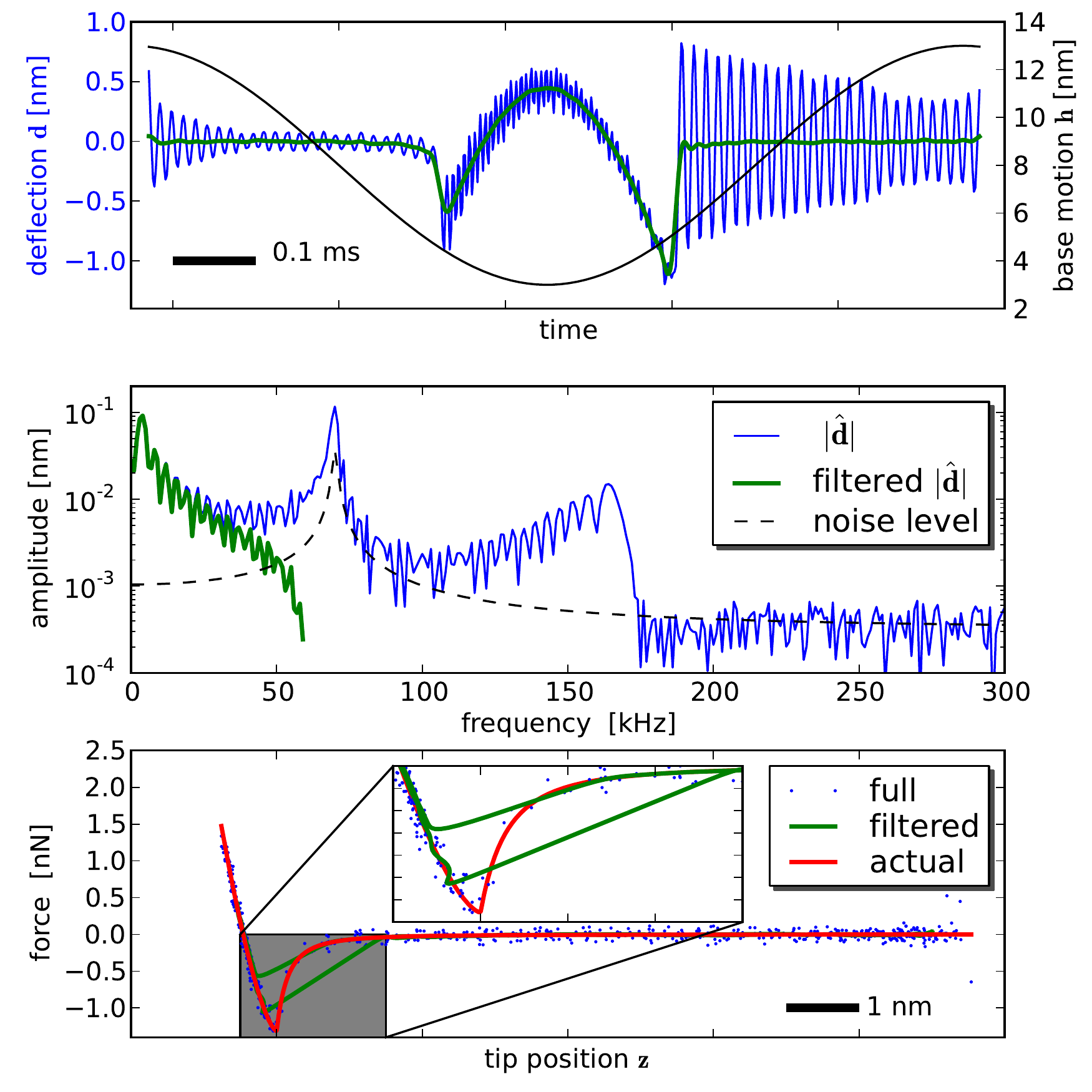}
\caption{Simulation of steady-state dynamics of a relatively soft cantilever, when driven by a sinusoidal base motion with frequency 1kHz.  The tip-surface forces acting on this soft cantilever result in motion spread over a wide band where the cantilever transfer function is strongly frequency dependant.  It is not possible to reconstruct the tip-surface force by filtering out the cantilever oscillations.  Simulation parameters are: cantilever stiffness $k_c=1$N/m, resonance frequency $f_0=70$kHz, quality factor $Q=50$,  tip surface force given in eq.~(\ref{DMT}) with parameters given below the equation. }
\label{fast_force_curve_70kHz}
\end{center}
\end{figure}

Figure~\ref{fast_force_curve_70kHz} shows a simulation of the cantilever dynamics for this mode of operation with the same tip-surface force and cantilever stiffness as for fig.~\ref{static_force_curves}.  The cantilever resonance was assumed to have  $f_0$=70~kHz and $Q$=50.  Simulations were made by numerical integration of the nonlinear ordinary differential equation eq.~(\ref{equ_mot_2}) using CVODE, part of the Sundials suite of nonlinear solvers \cite{SUNDIALS}.     Care was taken to properly treat the numerical integration in the neighbourhood of the contact point of the piece-wise-defined force model.  To make the simulation more realistic we added noise to the simulated deflection signal.    As explained in section \ref{sec:Sensitivity_Calibration}, the noise consisted of two contributions: a frequency-independent detector noise corresponding to an equivalent RMS deflection noise amplitude of 10~fm/$\sqrt{\text{Hz}}$; and a frequency-independent noise force with amplitude 23~fN/$\sqrt{\text{Hz}}$.  The force noise is identical to that of fig.~\ref{cantilever_noise}, but the detector noise level is a factor of 15 lower.  We choose this lower value of detector noise as it is achievable in the latest generation of commercial AFM's.  The noise was added to the deflection signal after simulation of the noise-free nonlinear dynamics, so our simulation neglects the very small nonlinear response due to the noise force.   

During one cycle of the base motion, the simulated deflection signal (fig.~\ref{fast_force_curve_70kHz} top panel) shows rapid oscillations at the resonance frequency of the cantilever.  As the base approaches the surface there is a sudden downward deflection when the cantilever jumps to contact, followed by a broad hump where the cantilever is deflected in contact with the surface, and a sudden upward deflection at the pull-off point.   The pull-off event excites the free oscillation of the cantilever at $f_0$, which rings down but does not completely extinguish before the next jump-to-contact event.  The jump-to-contact also excites higher frequency oscillations characteristic of the tip being bound to the surface.  In this simulation these oscillations decay with the same damping coefficient as the free oscillations, $\gamma_0$.

The middle panel of fig.~\ref{fast_force_curve_70kHz} shows the DFT of the deflection signal.  The DFT was taken over exactly one period of the drive signal, so each point of the DFT corresponds to a harmonic of the drive frequency, $f_\mathrm{drive}=1$~kHz.   In the spectrum one can see the free resonance frequency $f_0=70$~kHz, and a broader peak just below 175~kHz, containing jump-to-contact, pull-off, and oscillations in contact.  Above 175~kHz the deflection signal is essentially detector noise.

The bottom panel of fig.~\ref{fast_force_curve_70kHz} shows the reconstruction of the tip-surface force from the spectrum using eq.~(\ref{inversion_1}).   If we use the full response spectrum (blue line in the middle panel) the tip-surface force can be reconstructed as shown by the  blue dots in the bottom panel of fig.~\ref{fast_force_curve_70kHz}.  This inversion quite faithfully reproduces the actual tip-surface force used in the simulation (red  curve) albeit with noise.   We note that the reconstruction of the attractive part of the force curve, where the static force measurement method became unstable, only contains a few data points.  It is not possible to better resolve this part of the force curve simply by faster sampling of the motion.  One can see in the spectrum that we measure only noise at frequencies above 175~kHz.  The measurement is not limited by the bandwidth of data acquisition, but rather by the mechanical bandwidth of the force transducer.

%This is the result of the cantilever stiffness being much  smaller than the force gradients in the attractive region.  In this region the tip-surface force is rapidly accelerating the cantilever, giving rise to high-frequency components of the deflection signal.  One might argue that we should only sample faster in order to resolve this motion.  However, as apparent from the spectrum, there is only noise at frequencies above 175 kHz so faster sampling will only bring more noise in to the measurement.  The measurement is limited by the mechanical bandwidth of the force transducer, not the bandwidth of data acquisition.  

Here we should emphasize that in order to correctly reconstruct the tip-surface force, it is essential to capture the frequency components of the motion around resonance and above resonance, and it is necessary know the frequency dependence of the linear transfer function of the cantilever.  Without the former we can not reconstruct the tip-surface force near contact which cause rapid acceleration.  Without the later we can not separate the inertial, free damping and cantilever restoring forces from the tip-surface force, in their combined effect on cantilever motion.    One should not simply filter out the cantilever oscillations as demonstrated by the green curves in fig.~\ref{fast_force_curve_70kHz}.  Here we have applied a Hann filter with a cut-off frequency just below resonance.  This filter nicely removes the 'ringing' in the deflection signal (green curve, top panel of fig.~\ref{fast_force_curve_70kHz}) while retaining a dip feature at jump-to-contact, and deeper dip before pull-off.  A reconstruction of the tip-surface force from this filtered motion miss-represents the actual tip-surface force in the contact region completely (green curve, bottom panel fig.~\ref{fast_force_curve_70kHz}).  The filtered force curve is some kind of approximately quasi-static force curve with rounded jump trajectories, and one should not interpret the different branches on approach and retract as a measurement of the tip-surface interaction.  

\begin{figure}[t]
\begin{center}
\includegraphics[width=12cm]{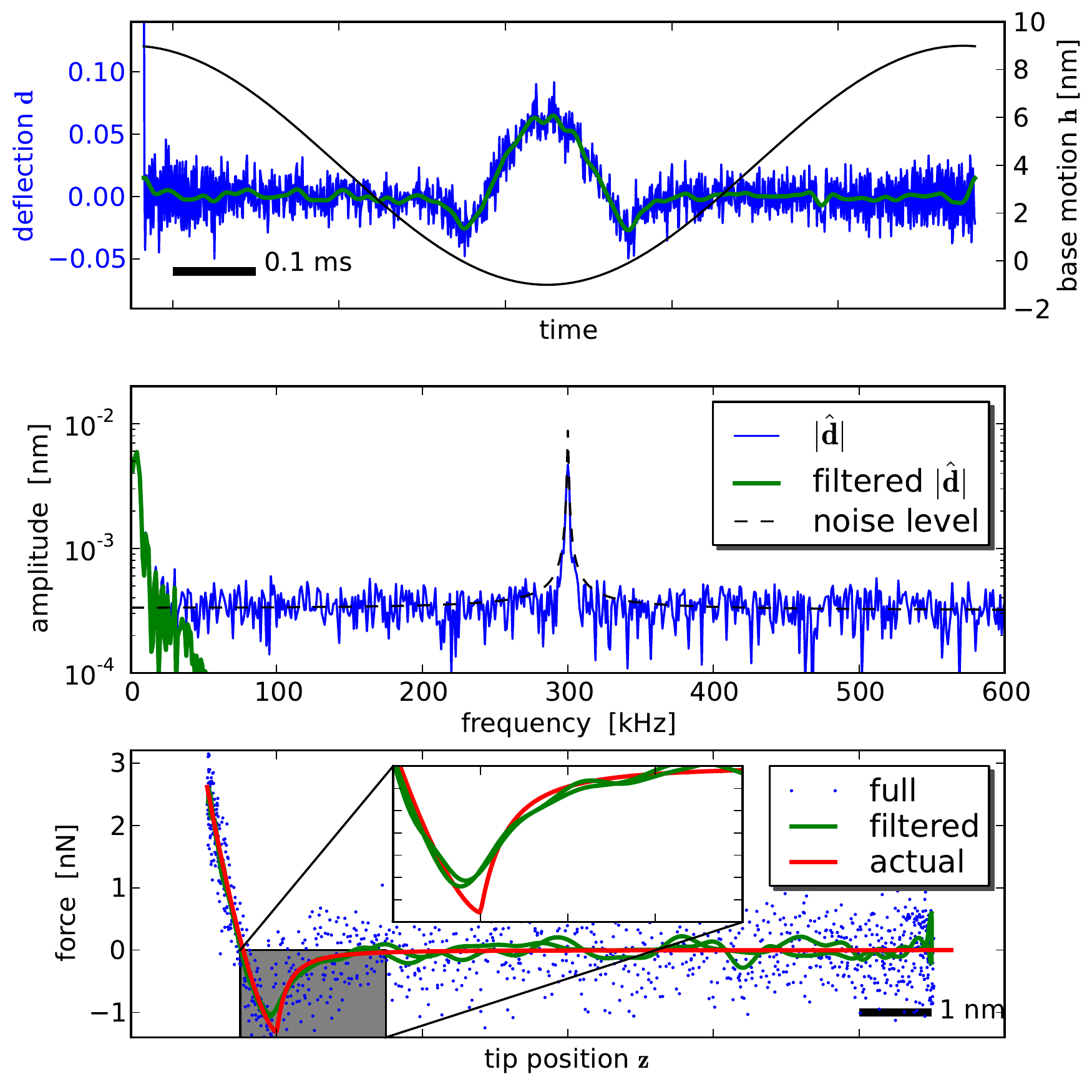}
\caption{Simulation of steady-state dynamics with the same tip-surface force as in fig.~\ref{fast_force_curve_70kHz}, but with a stiffer cantilever driven on resonance.  For this case, all significant motion occurs well below resonance and the force measurement is essentially quasi-static.  The stiffer cantilever gives poorer signal-to-noise ratio at these low frequencies.  Simulation parameters are: cantilever stiffness  $k_c=40$N/m, resonance frequency $f_0=300$kHz, quality factor $Q=468$,  tip surface force given in eq.~(\ref{DMT}) with parameters below the equation. }
\label{fast_force_curve_300kHz}
\end{center}
\end{figure}

The only solution to the problem of measuring the strongly attractive tip-surface force is to use a stiffer cantilever.  Increasing the stiffness for a fixed mass will increase the resonance frequency of the cantilever, thereby increasing its bandwidth as a force transducer.  To demonstrate this we preformed a simulation identical to that of fig.~\ref{fast_force_curve_70kHz} in all respects, differing only in the cantilever which had stiffness $k_c=40$~N/m.  The resonance was taken to have $f_0=300$~kHz, and $Q=468$, so that the noise force was the same $22$~fN/$\sqrt{\mathrm{Hz}}$.  Figure~\ref{fast_force_curve_300kHz} shows the result of this simulation.  The level of the deflection signal is smaller and the collision with the surface does not excite ringing above the noise level.  The jump-to-contact and pull-off events give the same change in the deflection signal (apart from the noise).  The high frequency part of the spectrum above $50$~kHz is only noise, so reconstruction of the tip-surface force from the full spectrum gives a rather noisy force curve (blue dots, bottom panel), this time however with more data in the contact region.   If we apply the same filter with a cut-frequency of $60$~kHz, we see that the reconstructed force curve more faithfully reproduces the tip-surface force in the contact region.  However the force curve suffers from rather much noise in the non-contact region and has difficulties reproducing the sharp kink at the contact point.  

Increasing the stiffness caused the frequency content of the motion to occur well below resonance where we can assume $\mathbf{\hat{G}} \simeq \mathbb{1}$.  In this case, the fast force curve is actually equivalent to rapidly taken quasi-static force curves.  Comparison of the two simulations presented in this section demonstrates that in order to reconstruct tip-surface force from motion, it is necessary to know the transfer function $\mathbf{\hat{G}}$ over the entire frequency band where there is significant motion.

\section{Force measurement near resonance}\label{sec:force_meas_resonance}

When the frequency content of the motion is far from a resonance, one looses the advantage of the increased signal due to the large transfer gain of a high quality factor resonance ($\vert \hat{G}(\omega_0) \vert = Q$).  Several methods have been devised which exploit this increased sensitivity.  In one particularly popular method known as frequency modulation AFM, the cantilever is driven with one pure tone in a phase-locked loop (PLL).  The PLL adjusts the drive frequency so that the phase difference between the drive and response is kept at a fixed value \cite{Albrecht:FMAFM:91}.  A force-distance curve can be generated by analysing the frequency shift while slowly moving the base toward and away from the surface \cite{Giessibl:FMAFMreview:03}\cite{Sader:QuantitiveForceMeasurement:05}.  While  high density force-volume maps have been made using this method \cite{Albers:3DdataAcquisition:09}\cite{Gross:ChemStructureAFM:09}, they are quite time consuming to acquire.  

Some measurement methods have been devised which excite the cantilever and measure at more than one resonance simultaneously \cite{martinez:2mode_afm:06}\cite{proksch:dual_acmode:06}\cite{Solares:TripleFreq:10}\cite{Martin:ProteinBimodal:11}.  While interesting images can be made with these methods, monitoring response at two or three frequencies contains too little information to be able to fully reconstruct the force.  Furthermore, the use of multiple eigenmodes complicates the dynamics considerably, making the force reconstruction problem and calibration much more difficult.  Non-resonant, multi-frequency methods have also been devised which use one drive tone and measure the response at many harmonics (integer multiples) of the drive tone \cite{Legleiter:SPAMinFluids:06}\cite{sahin:torsional_cantilever:07}\cite{stark:InvertingDynForceMic_02}.   When driving at resonance the response at these harmonics is effectively filtered out by the decaying transfer function ($G(\omega) \sim 1/\omega^2$ for $\omega > \omega_0$).  In order to better read out response at higher harmonics additional eigenmodes of the cantilever have been used \cite{sahin:torsional_cantilever:07}\cite{Sarioglu:TappingModeForceCurves:12}.    Here again, the use of multiple eigenmodes greatly complicates the force transducers dynamics and calibration.   Finally, some techniques have considered two drive tones near a resonance with frequency modulation AFM \cite{Rodriguez:DART:07} and continuous band excitation near resonance \cite{Jesse:BandExcitation:07}, where the response was analysed within the context of linear dynamics.

\section{Intermodulation AFM}\label{sec:ImAFM}

The difficulties associated with broad band response caused by the nonlinear tip-surface force can be circumvented by exploiting the phenomena of frequency mixing, or intermodulation, which occurs in nonlinear systems that are driven by more than one pure tone.  In this case, not only harmonics of the drive frequencies are generated by the nonlinearity, but also intermodulation products, which occur at integer linear combinations of the drive frequencies $\omega_1, \omega_2, \omega_3 \dots$,
\begin{equation}
\omega_{\rm IMP} = n_1 \omega_1 + n_2 \omega_2 + n_3 \omega_3 + \dots
\end{equation}
where $n_1, n_2, n_3 \dots$ are integers.  The order of an intermodulation product is given by $\vert n_1 \vert + \vert n_2 \vert + \vert n_3 \vert + \dots$.  With proper choice of the drive frequencies, we can induce a response where many intermodulation products of high order occur near resonance.  This will ensure that information about strongly non-linear terms in the oscillator equation of motion can be measured with good SNR.  

When measuring intermodulation products it is necessary that the drive frequencies be taken from a discrete set of tones which are the integer multiples of $\Delta \omega$.  In this case, all intermodulation products will occur at frequencies that are contained in the DFT and one can therefore extract both the amplitude and phase of the response at intermodulation frequencies by calculating Fourier sums as in eq.~\ref{DFT}.  This condition on the drive frequencies simply means that the applied drive waveform is periodic in the time window $T=2\pi / \Delta \omega $.  The fundamental assumption we make in intermodulation AFM is that the weakly nonlinear motion will be periodic, with the same period as the drive waveform.  We note that for strongly nonlinear motion, where bifurcations to sub-harmonic response occurs {\it en route} to chaos, our analysis based on the assumption of periodic response could break down.   

Perhaps the most simple way to preform intermodulation AFM is to choose two drive tones, both integer multiples of a base tone, which are approximately centred around the cantilever resonance and separated in frequency by the linewidth of the resonance $\gamma_0$ \cite{Platz:ImAFM:08}.  When the drive amplitudes are adjusted to give equal amplitude response at each frequency, the free oscillation of the cantilever will form a beating waveform in the time domain, as displayed in the top panel of fig.~\ref{intermodulation_force_curve}.  Here the engaged response is simulated for the same van der Waals - DMT nonlinear force used for all previous simulations, with the stiffer cantilever, $k_c$=40 N/m, used for fig.~\ref{fast_force_curve_300kHz}.   The response waveform of the cantilever engaging the surface (blue curve) has an envelope function slightly different from of the perfect beat, which can hardly be seen in the time-domain data.  The spectrum however shows response not only at the two drive frequencies, but also several new peaks clustered around resonance, which are the intermodulation products of the two drive tones (inset of fig.~\ref{intermodulation_force_curve}, middle panel).  Note that a cluster of intermodulation products can also be seen just above the noise floor, around the second harmonic of the center frequency, $2\bar{\omega}=(\omega_1+\omega_2)$.  The amplitude of these peaks is however greatly diminished due to the fact that they occur off resonance.  

The intermodulation spectrum near resonance represents essentially all the signal that is possible to measure above the detector noise floor.  This data is only a {\em partial} spectrum of the actual motion.  Detector noise and limited bandwidth of the force transducer inhibit our ability to measure the entire spectrum.  If  the entire spectrum could be measured in full detail, the reconstruction of force would be given by eq.~\ref{inversion_1}.   The essential question then becomes:  To what extent can we reconstruct the nonlinear force from analysis of the partial spectrum?  The answer is that we can do an excellent job in spite of the narrow detection bandwidth because the motion spectrum contains many intermodulation products to rather high order.  These intermodulation products contain much information about the nonlinearity that generated them.

\begin{figure}[t]
\begin{center}
\includegraphics[width=12cm]{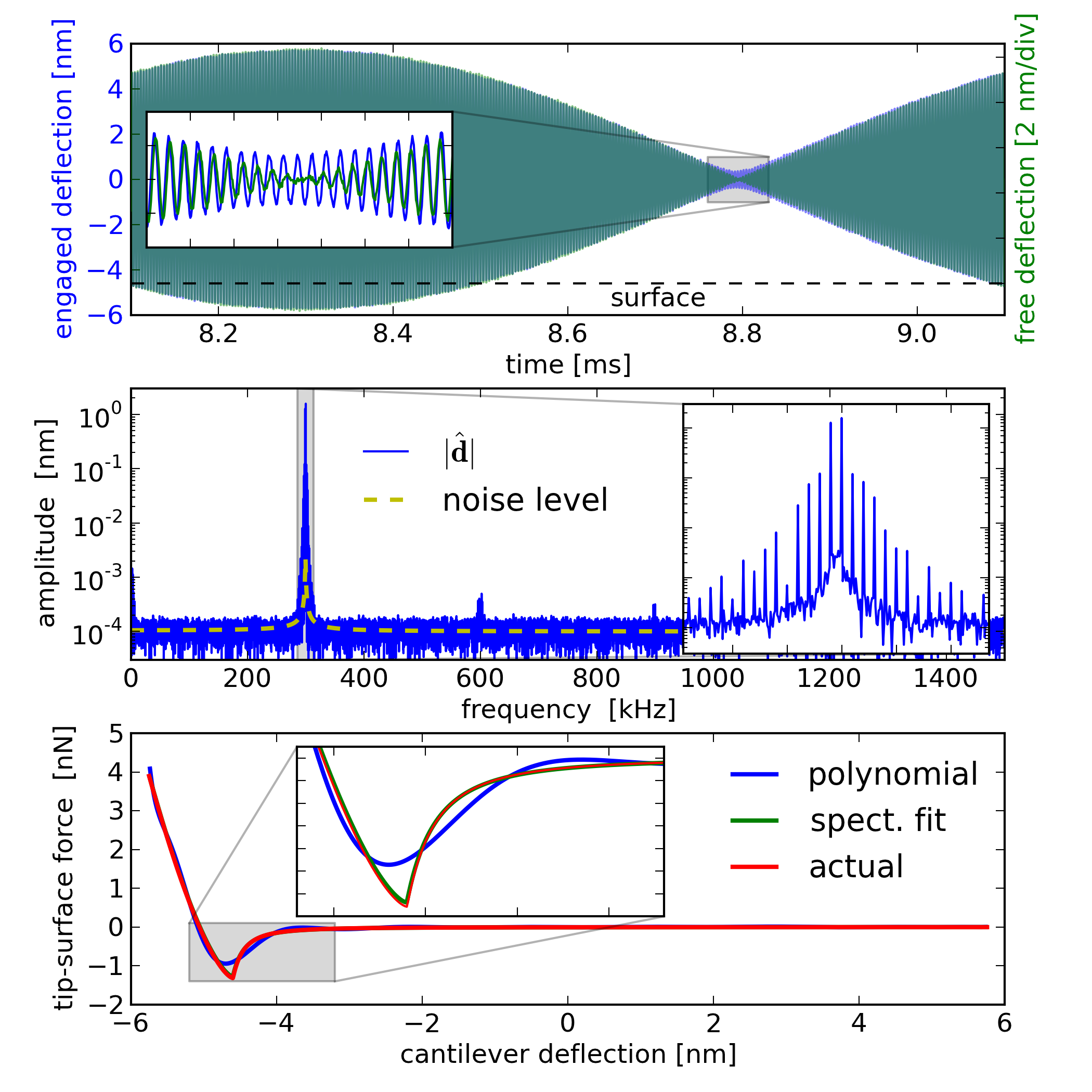}
\caption{Simulation of intermodulation AFM for the same cantilever and tip-surface force parameters as that of fig.~\ref{fast_force_curve_300kHz}, but now driven with two closely spaced tones centred on resonance.  The free cantilever motion in the time domain is a beating waveform.  The engaged motion has a slightly distorted beat envelope which appears in the frequency domain as intermodulation products, or mixing products of the two drive tones.  The inset in the middle panel shows a detail of  the engaged response spectrum near resonance.  Only the amplitude is displayed, but the phase of each intermodulation product is also measured.  Through analysis of this intermodulation spectrum we can reconstruct a polynomial approximation of the tip-surface force, or reconstruct the parameters of the force model used in the simulation.}
\label{intermodulation_force_curve}
\end{center}
\end{figure}

\section{Reconstruction from a partial spectrum}\label{sec:Reconstruction}

Suppose we have a partial spectrum $ \mathbf{\hat{d}}^{\mathrm{(part)}}$, which consists of $M$ components $\lbrace \mathrm{\hat{d}}_{k_1},  \mathrm{\hat{d}}_{k_2}, \dots, \mathrm{\hat{d}}_{k_M} \rbrace$ .  We use this partial spectrum of engaged motion, the spectrum of free motion, and the calibrated transfer function to reconstruct the tip motion in the lab frame,
\begin{equation}
\mathbf{z} =   \mathcal{F}^{-1} \left[ \mathbf{\hat{d}}^{\mathrm{(part)}} + ( \mathbf{\hat{G}} -  \mathbb{\hat{1}} )^{-1}  \mathbf{\hat{d}}^\mathrm{(free)} \right]
\label{partial_motion}
\end{equation}
In order that this be an accurate representation of the actual motion, it is important that our partial spectrum contains all significant peaks near resonance, and all peaks near higher harmonics of $\bar{\omega}$ are negligible.   Here we note that because we are driving near resonance, we  an easily measure both $\mathbf{\hat{d}}^\mathrm{(part)}$ and $\mathbf{\hat{d}}^{\rm (free)}$  from the same drive $\mathbf{\hat{h}}$, so it is possible to reconstruct the motion $z(t)$ from analysis of  {\em only deflection signals}.  There is no need for an independent measurement of the base motion $h(t)$.  This means that the calibration of all measurements needed to generate $\mathrm{f}(z)$ are traceable to one thermal noise measurement which is easily performed before scanning, and can be frequently checked during a scan session.  One does not rely in any way on the scanner calibration.  For this reason, we will plot force directly in terms of the cantilever deflection in fig.~\ref{intermodulation_force_curve}. 

One approach to reconstructing the nonlinear force is to assume a particular nonlinear force model described by a function that contains $P$ parameters $g_i$.   We then define an error function which is the difference between the nonlinear force at the frequencies measured, and that calculated from the measured motion $\mathbf{z}$,
\begin{equation}
\mathbf{\hat{e}} = k_c \mathbf{\hat{G}}^{-1} 
\left(   \mathbf{\hat{d}}^{\mathrm{(part)}} - 
\mathbf{\hat{d}}^{\mathrm{(free)}}  \right) -
\mathcal{F} \left[ \mathrm{f}( g_1,g_1,\dots,g_P; \mathbf{z}) \right]
\end{equation}
Using numerical optimization we adjust the parameters of the model function so as to minimize  the $M$ components of $\mathbf{\hat{e}}$ where we have data with good SNR \cite{Forchheimer:SpectFit:12}.  

 Figure~\ref{intermodulation_force_curve} shows the result of a least-square minimization of the sum $\sum_{k=1}^M(\mathrm{Re[\hat{e}}_k])^2+(\mathrm{Im[\hat{e}}_k])^2$ where the DMT model was assumed.  The red curve is the actual force, and the green curve is the reconstructed force found by adjusting $P=4$ parameters of the DMT model to fit the noisy partial spectrum consisting of $M=28$ components around resonance.  We do not adjust the tip radius because the DMT model does not depend on $R$ in an unambiguous way.   We can see that in spite of the noise, the reconstruction is nearly perfect, missing the actual force only slightly near the contact point where the nonlinearity changes rapidly.  
 
 This reconstruction from simulated data with noise demonstrates that the partial spectrum does contain enough information with good enough SNR to fully reconstruct the force.  It is however rather artificial in that we are starting with a model function which has the exact same form as the nonlinearity which actually caused the motion.   In a real experiment we do not know what model best describes the true force.  The spectral fitting method can therefore be dangerous to use without reference to the quality of the fit, because it will often converge to some parameters that describe a noiseless, ideal force curve having the shape of the model function.   This type of analysis should therefore be complimented with an independent method, to be sure that the assumed model is physically meaningful.  Nevertheless, if one does have a good model, this spectral fitting method represents a way to optimally extract the model parameters directly from the measured data.  

The ultimate goal of quantitative dynamic AFM to devise a method to reconstruct force from motion, which does not depend on an assumed force model.  This reconstruction is however an ill-posed problem, and certain assumptions must be made in order to determine tip-surface force from a partial representation of the motion.    The art of finding a good reconstruction method is to make assumptions which are physically well motivated, but minimally constrain the possible nonlinear functions that can be achieved.   For example, one may assume that the nonlinear tip-surface force can be well approximated by a polynomial in $z$ of degree $P$ on the finite interval of $z$ over which the cantilever oscillates \cite{Hutter:Reconstructing:2010}.
\begin{equation}
\mathrm{f}(z) \simeq k_{\rm c} \sum_{j=0}^{P-1} g_j z ^j  
\textrm{     for     }z \in \lbrace z_\mathrm{min} , z_\mathrm{max}  \rbrace
\label{polynomial_force}
\end{equation}
Since this force is linear in the polynomial coefficients, its DFT can be written as a matrix equation,
\begin{equation}
\mathbf{\hat{f}} = k_\mathrm{c} \mathbf{\hat{H}} \mathbf{g} 
\label{fhat_k}
\end{equation}
where the components of the vector $\mathbf{g}$ are the polynomial coefficients, and the matrix $\mathbf{\hat{H}} $ is a $M \times P$  matrix constructed from the DFT of $z^j$.  
\begin{equation}
\hat{H}_{kj} =   \mathcal{F}_k \left[ z ^j \right]
\label{H_matrix}
\end{equation}
The equation for the engaged response, eq.~(\ref{engaged_spectrum}),  then becomes 
\begin{equation}
\mathbf{\hat{d}} = \mathbf{\hat{d}}^{\rm (free)} + \mathbf{\hat{G}} \mathbf{\hat{H}} \mathbf{g} 
\label{poly_d_hat}
\end{equation}
Taking only the measured intermodulation frequencies, we have a system of $M$ equations which can be inverted to solve for the polynomial coefficients,
\begin{equation}
\mathbf{g}=\mathbf{\hat{H}}^{+} \mathbf{\hat{G}}^{-1} (\mathbf{\hat{d}}^{\rm (part)} - \mathbf{\hat{d}}^{\rm free})
\label{poly_inversion}
\end{equation}
where the matrix $\mathbf{\hat{H}}^+$ is the pseudo inverse of $\mathbf{\hat{H}}$.  The greater the number $M$ of Fourier coefficients in the partial spectrum $\mathbf{\hat{d}}^\text{(part)}$ which have good SNR, the larger the number $P$ of polynomial coefficients that can be determined.  When $M>P$ the system of equations (\ref{poly_inversion}) becomes overdetermined and in this case the pseudo inverse finds the polynomial coefficients which best fit the measured data in a least-squared sense.   

Applying this polynomial inversion to a partial spectrum such as that shown in middle panel inset of fig.~\ref{intermodulation_force_curve} will however not work directly.  The drive scheme used to generate this spectrum produces only intermodulation products of odd order near resonance, where the dominant response comes from coefficients of $z^j$ with odd $j$ (first order response in perturbation theory).  The partial spectrum of odd intermodulation products contains very little information about the even polynomial coefficients.   We can however easily fix this problem by applying an additional, physically well-motivated constraint to our reconstruction algorithm.  We know that the tip-surface interaction only occurs near one turning point of the cantilever oscillation.  The constraint $\mathrm{f}=0$ for $d>0$ can then be used to determine the even $g$'s, given that the odd $g$'s are determined as described above.  

The result of this reconstruction method is shown in the bottom panel of fig.~\ref{intermodulation_force_curve} with the blue line, where $P=19$ polynomial coefficients were determined from the intermodulation spectrum  consisting of $M=28$ components.  Over the entire interval of oscillation, the polynomial captures the shape of the DMT force curve quite well.  The zoom of the contact region shows that the polynomial curve wiggles around the actual force curve.  These wiggles are not due to the noise in the spectral data, they are rather the nature of a polynomial curve which is trying to approximate a function that is flat everywhere on the interval, except for one end where it changes rapidly and has a very sharp kink.   Here we should point out that the piece-wise-defined force model used in the simulations is a very demanding nonlinearty to reconstruct.  In real experiments we expect that actual forces will not have discontinuous force gradients as with this model function.  Thus, the polynomial method can be quite good with experimental data.  Using simulations such as these, we have determined that the polynomial generated by the inversion algorithm described above is very close to the optimal interpolation polynomial that approximates the nonlinear function.  No significant improvement on this general method can be expected by assuming a different type of polynomial, for example a Chebychev polynomial \cite{Durig:InteractionSensing:00}.    

The polynomial reconstruction has the distinct advantage that it captures the general shape of the force curve without assuming a particular force model.  The equations are linear in the polynomial coefficients and the matrix inversion gives us a {\em unique} set of coefficients.  The key step in the polynomial inversion is building the matrix $\mathbf{\hat{H}}$ and inverting it.  The columns of $\mathbf{\hat{H}}$, eq.(\ref{H_matrix}), can be built up by recursive multiplication of the motion, eq.(\ref{partial_motion}), followed by a DFT. The DFT's can be done using the FFT alogrithm, and the matrix calculations are not very large, so the entire reconstruction algorithm is quite fast.  The non-optimized Numpy code used to invert the data for  fig.~\ref{intermodulation_force_curve} took less than 82 ms to run on a laptop.  An AFM which is equipped to rapidly capture the intermodulation spectrum at each pixel  allows the operator to get immediate feedback in the form of a calibrated force curve at any chosen image pixel \cite{Tholen:IMlockin:10}\cite{Intermodulation-Products}.  Depending on the shape of the polynomial curve, the operator can switch inversion methods and quantitatively determine the physical parameters of an arbitrary force model which best fit the measured data.

\section{Conservative and dissipative forces}\label{sec:conservative_dissipative}

Our analysis of dynamic force measurements thus far has been based on simulations of the nonlinear dynamics of a cantilever with conservative tip-surface interactions.  In real experiments we can expect that dissipative processes will be present in the tip-surface interaction, due to a non-elastic response of the material under deformation, the rearrangement of adsorbate molecules on the surface, or the making and breaking of chemical bonds when the tip makes and brakes contact with the surface.  Many possible microscopic irreversible processes can give rise to dissipation, or energy loss of the oscillating cantilever.  These dissipative interactions do not exist when the cantilever and surface forces are in static equilibrium.  Dissipative interactions can only be probed with dynamic modes of AFM.  Understanding the origin of and reconstructing dissipative interactions is one of the grand challenges of quantitiative AFM\cite{Sader:QuantitiveForceMeasurement:05}\cite{Durig:InteractionSensing:00}\cite{Raman:InvertingAmplitudePhase:08}.

\begin{figure}[t]
\begin{center}
\includegraphics[width=12cm]{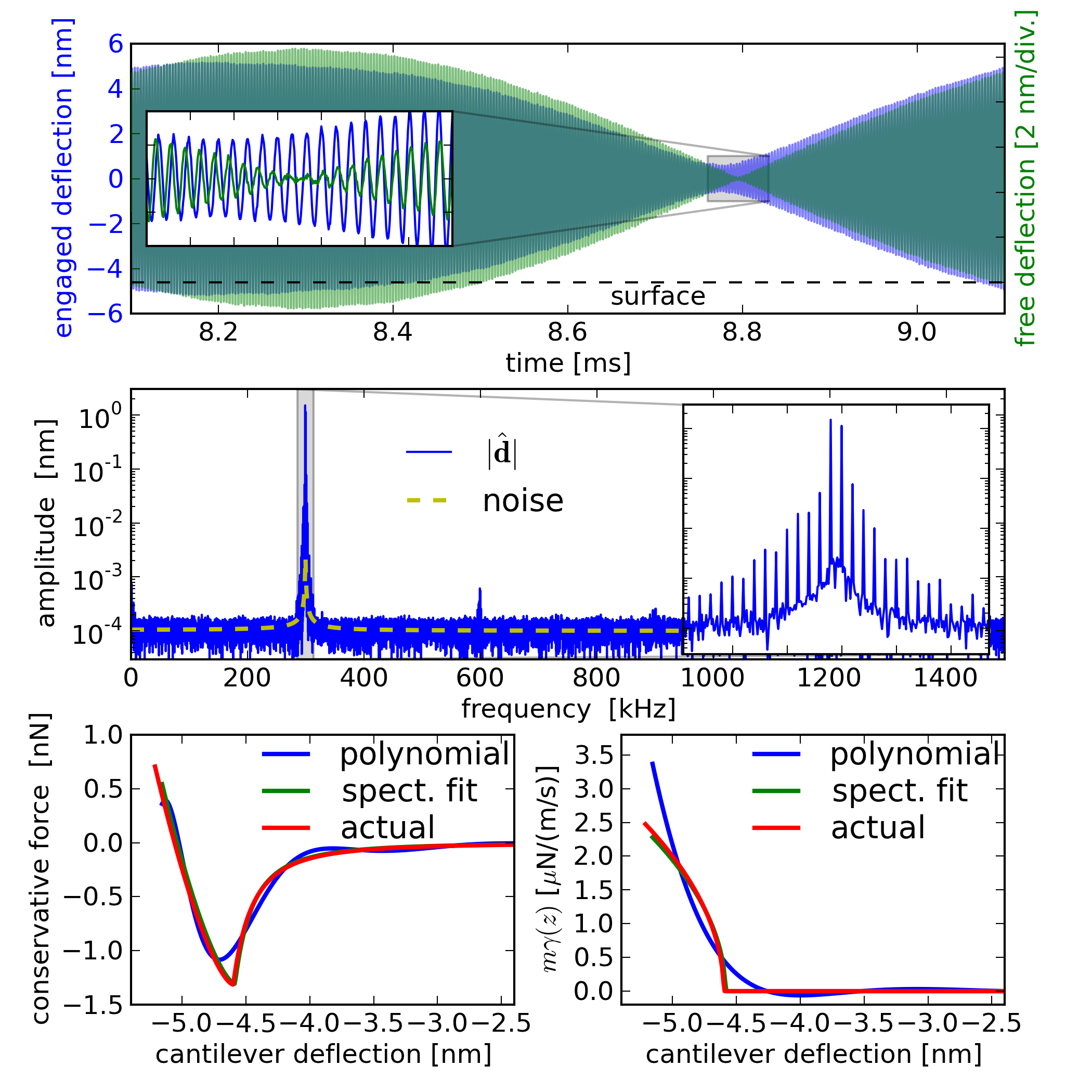}
\caption{Simulation of intermodulation AFM with the same parameters as in fig.~\ref{intermodulation_force_curve}, but this time with an additional Kelvin-Voigt surface damping force.  The dissipative forces causes greater distortion of the free beating waveform.  The intermodulation spectrum contains the information needed to reconstruct both the conservative force and the damping function. The reconstruction of the conservative force is completely uneffected by the dissipative interaction.  Due to energy loss of the oscillator, the maximum tip-surface force has decreased in comparison with fig.~\ref{intermodulation_force_curve}.  }
\label{intermodulation_disipative_force_curve}
\end{center}
\end{figure}

The intermodulation spectrum does contain information about dissipative tip-surface interactions, information which was not used in the reconstruction algorithms described previously for conservative forces.   Within the context of the single eigenmode model which has only two degrees of freedom, $z$ and $\dot{z}$,  dissipation can be described by a velocity-dependant tip-surface force.  The determination of this force becomes the task of reconstructing an arbitrary nonlinear function $\mathrm{F}(z,\dot{z})$ from the measurable partial spectrum of intermodulation products.  Let us assume a simple model, where the total tip-surface force can be described by a conservative force, plus a position-dependant viscous damping force.  
\begin{equation}
\mathrm{F}(z,\dot{z}) = \mathrm{f}(z) +  m\gamma(z)\dot{z}
\end{equation}
Approximating the damping function by a polynomial in $z$,
\begin{equation}
\gamma(z) \simeq \sum_{j=1}^{P} g^\mathrm{(dis)}z^j
\end{equation}
and using the same polynomial approximation for the conservative force with coefficients $\mathbf{g}^{\rm (con)}$,  we arrive at an equation of motion similar to eq.(\ref{poly_d_hat}).
\begin{equation}
\mathbf{\hat{d}}=\mathbf{\hat{d}}^{{\rm (free)}}+\mathbf{\hat{G}\left\{ \mathbf{\hat{H}},\mathbf{\hat{I}}\right\} }\left\{ \begin{array}{c}
\mathbf{g^{\mathrm{(con)}}}\\
\mathbf{g^{\mathrm{(dis)}}}
\end{array}\right\} 
\label{poly_d_hat2}
\end{equation}
where the braces denote the horizontal and vertical concatenation of the matrices and vectors respectively.  The matrix $\mathbf{\hat{I}}$ is formed in a way analogous to the matrix $\mathbf{\hat{H}}$, 
\begin{equation}
\hat{I}_{kj} =\omega_0^{-2} \mathcal{F}_k \left[ \dot{z} z ^j \right] 
\end{equation}
Multiplication in the time domain is a convolution in the frequency domain, so the matrix $\mathbf{\hat{I}}$ can be calculated by convolution of the velocity vector $\hat{\dot{z}}_k=ik\Delta\omega \hat{z}_k$  with the columns of $\mathbf{\hat{H}}$.
The vector of polynomial coefficients describing the conservative force, $\mathbf{g^{\mathrm{(con)}}}$ and position dependant viscosity, $\mathbf{g^{\mathrm{(dis)}}}$ are then found by inverting eq.~(\ref{poly_d_hat2}),
\begin{equation}
\mathbf{\left\{ \begin{array}{c}
\mathbf{g^{\mathrm{(con)}}}\\
\mathbf{g^{\mathrm{(dis)}}}
\end{array}\right\} }=\mathbf{\left\{ \mathbf{\hat{H}},\mathbf{\hat{I}}\right\} ^{+}G^{-1}}\left(\mathbf{\hat{d}}-\mathbf{\hat{d}}^{{\rm (free)}}\right)
\end{equation}

Dissipative processes can also be incorporated in a model function describing the tip-surface force.   If the dissipative process is parametrized, then the spectral fitting method can extract the parameter values which best fit the data.  Because dissipation can be due to so many different factors, it is difficult to argue that one particular model is more valid than another.  Nevertheless, one very attractive feature  of the spectral fitting method is that it can accommodate essentially any interaction model, such as double-value functions and forces which turn on and off instantaneously depending on the motion history and the tip position and velocity.  As long as the interaction can be programmed, numerical optimization can be preformed.  

To demonstrate force reconstruction including dissipative forces, we simulate the nonlinear dynamics using the van der Waals - DMT - Kelvin-Voigt model for the tip-surface force \cite{melcher:061301},
\begin{equation}
\mathrm{f}(z)=\begin{cases}
\begin{array}{c}
-\frac{HR}{6z^2}\\
-\frac{HR}{6a_{0}^2}+\frac{4}{3}E^{*}\sqrt{R(a_0-z)^{3}} -\eta \dot{z}\sqrt{R(a_0-z)}
\end{array} & \begin{array}{c}
z \geq a_{0}\\
z<a_{0}
\end{array}\end{cases}
\label{DMT_KV}
\end{equation}
which is a simple modification to the previous model eq.~(\ref{DMT}) to include a viscous damping described by the parameter $\eta$, that increases as the square root of the penetration in the contact region. 

Figure \ref{intermodulation_disipative_force_curve} shows the result of a simulation using this force.  The simulation is identical to that done for fig.~\ref{intermodulation_force_curve}, the only difference being that we added the dissipative force with $\eta=1000$Pa-s.  One can see a more significant change between the free oscillation and the engaged oscillation in the time domain (top panel).  The frequency domain (middle panel) shows that nearly all motion is still contained in the narrow band around resonance, where several intermodulation products can be resolved.  The reconstruction of the conservative force and damping function are shown in the bottom panels of fig.~\ref{intermodulation_disipative_force_curve} where we see that the spectral fitting method can accurately recover  the nonlinear functions used to generate the simulated data.  We note that the ability to recover the conservative force is completely unaffected by the addition of the dissipative force in the simulation.  The polynomial damping function reconstructed from the intermodulation spectrum is able to capture the rapid turn-on of the dissipative interaction at the very end of the oscillation, but a polynomial can not fully approximate actual damping function used in the simulation.   The inaccuracy in the reconstruction of the damping curve with both methods is greatest at the endpoint of the oscillation.   This turning point, where the velocity goes to zero, is a singular point for any viscous damping model, so we can not expect to measure dissipation accurately in this region of the oscillation.  Nevertheless, this simulation clearly demonstrates how intermodulation AFM enables one to determine the region of the oscillation cycle where the dissipation is occurring. 

\section{Summary and conclusion}

In this paper we looked at the problem of determining tip-surface force from measurement and analysis of cantilever motion.  In contrast to quasi-static measurement, dynamic measurement of the tip-surface force must properly account for the  contributions of inertia and damping on the cantilever motion.  We focused on the single eigenmode model as this is an excellent description of the cantilevers free linear dynamics in a frequency band surrounding a resonance.  Our approach to understanding the engaged dynamics was to consider the full effect of the nonlinear tip-surface force, without approximating an effectively linear oscillator dynamics.   Assuming that the steady-state response of the nonlinear oscillator will have the same periodicity as the drive signal, we considered a discrete frequency domain analysis of the motion.  The effect of the nonlinearity on multiple drive tones contained in this discrete domain, was the generation intermodulation products that are also contained in the discrete domain.  

A special case of Intermodulation AFM was simulated with a drive signal consisting of two pure tones close to resonance that generated a response with many odd-order intermodulation products near resonance.  Using this scheme we can measure the amplitude and phase of intermodulation products to rather high order, in a frequency band where sensitivity is highest and where accurate calibration can be preformed.  For these reasons, Intermodulation AFM stands out as being  a very accurate and sensitive way to extract the tip-surface force.  The information contained in the intermodulation spectrum allows one to reconstruct not only conservative, but also dissipative tip-surface forces.  In this respect, dynamic force measurements offer a wealth of new possibilities for quantitative AFM that go far beyond what one can do with static force measurements.  

In conclusion we would like to point out that the two methods of force reconstruction presented here, one based on numerical optimization with arbitrary force models and the other based on matrix inversion with a linear expansion of the force,  are not the only possible methods of reconstructing nonlinearity from an intermodulation spectrum.  One can look at intermodulation spectroscopy as technique for transposing information about the nonlinearity, to the narrow frequency band around resonance where it can be revealed above the detector noise floor.  Various schemes for driving the nonlinear oscillator can be envisioned,  which differently transpose information about the nonlinearity.  Once transposed, the intermodulation spectrum can be analyzed in a variety of different ways to extract the desired information from the measurable nonlinear response.  From this point of view the intermodulation measurement can be seen as a 'compression sensing technique' which efficiently extracts and stores relevant information from the detector signal, rejecting only noise.  This compression aspect of intermodulation AFM gives it great advantage over competing force measurement methods in that it enables storage of the complete raw data set for offline analysis.   Thus, different hypothesis can be tested on the same data set, and a much more detailed study of the tip-surface interaction at every image point can be made.  

\section*{Acknowledgements}
We gratefully acknowledge financial support form the Swedish Research Council (VR) and the Swedish Governmental Agency for Innovation Systems (VINNOVA).

\section*{References}

\bibliographystyle{unsrt}
%\bibliography{David}

\end{document}